# A Low Cost EEG Based BCI Prosthetic Using Motor Imagery

Daniel Elstob[1] & Emanuele Lindo Secco[1]

[1]Department of Mathematics & Computer Science, Liverpool Hope University, Liverpool, UK

## Abstract

*Brain Computer Interfaces (BCI) provide the opportunity to control external devices using the brain ElectroEncephaloGram (EEG) signals. In this paper we propose two software framework in order to control a 5 degree of freedom robotic and prosthetic hand. Results are presented where an Emotiv Cognitive Suite (i.e. the 1st framework) combined with an embedded software system (i.e. an open source Arduino board) is able to control the hand through character input associated with the taught actions of the suite. This system provides evidence of the feasibility of brain signals being a viable approach to controlling the chosen prosthetic. Results are then presented in the second framework. This latter one allowed for the training and classification of EEG signals for motor imagery tasks. When analysing the system, clear visual representations of the performance and accuracy are presented in the results using a confusion matrix, accuracy measurement and a feedback bar signifying signal strength. Experiments with various acquisition datasets were carried out and with a critical evaluation of the results given. Finally depending on the classification of the brain signal a Python script outputs the driving command to the Arduino to control the prosthetic. The proposed architecture performs overall good results for the design and implementation of economically convenient BCI and prosthesis.*

## 1. Introduction

With the current knowledge of how brain works, researchers are able to develop a wide range of applications that can improve life quality to those with muscular or motor-neuron disabilities using devices called Brain Computer Interfaces (BCI). BCI is defined as "a communication and/or control system that allows real-time interaction between the human brain and external devices" (Mak and Wolpaw, 2009).

In relation to history of BCIs, in 1973 Vidal was researching into "direct brain-computer communication" (Vidal, 1973) and "Real-time detection of brain events in EEG" (Vidal, 1977). Up until the 1990's research did not progress much at which point there was a breakthrough in the research with the pioneering research done by Wolpaw et al. who produced an alternate approach to a BCI system using ElectroEncephaloGrams (EEG, Wolpaw et al., 1990). Another communication approach by Wolpaw et al. was designed for a system with a cursor on the screen which the subject would control with thought (Wolpaw et al., 1991). In 1998 Farwell and Donchin designed another approach to use "the P300 component of the event-related brain potential (ERP)" (Farwell and Donchin, 1998) in order for individuals with motor related disabilities to have a form of communication. This system required the user to concentrate on screen filled with letters, each flashing at a different frequency, at which point the computer would, depending on the P300 signal received, detect the chosen letter. Levine et al. explored "a direct brain interface based on event-relate potential" (Levine et al, 2000) with this research presenting a system capable of equal accuracy as the other current communication systems.

Literature in the field has focused on developing a BCI that could assist or repair human cognitive functions such as restoring hand grasp (Pfurtscheller et al., 2003) or even augmentations like controllable prosthesis which "aim to provide a communication channel equivalent to "typing" on a computer" (Sajda et al., 2008). As a result of this the research and development of BCIs are largely towards neuro-prosthetics, prosthetics that aim towards the rehabilitation of patients; for instance Müller-Putz et al. showed that BCI can be used "for the control of neuro-prosthesis in patients with high spinal cord lesions" (Müller-Putz et al., 2005).

Ultimately the purpose of a BCI is to acquire the desired signals or intent of the user however these BCIs can be either invasive or non-invasive depending on the requirements of the user. An invasive BCI means that the system or device is integrated into the individual: for instance electrodes directly into the brain matter or partially integrated into the inner skull. Non-invasive can operate outside the individual's body (Smith and Delargy, 2005). Although this is well established field of research the real world applications for a non-invasive approach has focused towards the domains of virtual reality (Anatole et al., 2008) and gaming (Nijholt, 2009 and Rossini et al., 2009).

The awareness that EEG BCI can improve the daily lives of some - if not all - patients is a common idea. Höhne et al. presented "a BCI system designed to establish external control for severely motor-impaired patients within a very short time" (Höhne et al., 2014). This study included patients of various degrees of disabilities with two being in a locked-in state. This study showed how valuable and feasible a BCI system for these types of disabilities are as "within only six experimental sessions, three out of four patients were able to gain significant control over the BCI" (Höhne et al., 2014) and they further explain that their system could outperform some of the best assistive technologies these patients were using. Other researchers, such as Bougrain et al. performed studies on "decoding intracranial data recorded in the cortex of a monkey and replicates the associated movements on a JACO robotic arm by Kinova" (Bougrain et al., 2012). Qin et al. performed "classification of motor imagery for brain-computer interface applications, by means of source analysis of scalp-recorded EEGs" (Qin et al., 2004) and achieved a classification rate of around 80% from their subjects. Finally, Johnson researches into an "EEG based BCI that uses steady-state visual evoked potentials on a healthy individual to asynchronously control a 6-degree of freedom robotic arm through custom software in real-time with high accuracy (Johnson, [no date]). This study implemented a BCI for healthy human rather that for medical related reasons but uses a system that could be slow for a control system of external device such as a prosthetics due to the nature of Steady-State Visually Evoked Potentials (SSVEP) which requires the user to look at a symbol or an area of a screen that is flashing repeatedly causing a specific signal in the brain that will be recognised by the system and perform a task.

This work aims to design a working prosthetic hand that uses non-invasive EEG based motor imagery and provides the user the ability to control the prosthetic hand using motor imagery signals. To do this the acquired EEG signal will be processed and classified to produce the highest accuracy possible that will be used as an output for control. The purpose of this work is to use a consumer grade headset, namely an Emotiv Headset, to acquire raw EEG signals and process this data in order to identify and classify the intent of the user. This intent will then be used to control a robotic or prosthetic device. An OpenViBE platform will be used as a foundation to develop a customized interface in order to record and process signals, to classify the extracted and desired features and turn them into commands, finally providing a feedback system for those commands.

In order to achieve these targets, the following tasks have been broken down:

- Acquisition and recording of the EEG signals
- Filtering of EEG signals
- Training of a classifier
- Analysis of classification performance and accuracy
- Performing a real-time scenario with feedback
- Performing a real-time analysis
- User intent to control a prosthetic

Paragraph 2 introduces the methodology of the research and the flow of processing the signal in order to create the BCI system. Paragraph 3 will present the steps of implementing this project with the testing of the prosthetic with serial monitor inputs to Arduino. Paragraph 4 performs an evaluation of the results produced during the implementation and, finally, Paragraph 5 provides and overall conclusion of the project including the methodology used, the limitations and BCI illiteracy of users when using the systems, the robustness and the real world or clinical applicability of the created BCI systems finishing with a final conclusion.

## 2. MATERIALS & METHODS

In this paragraph the hardware and software will be presented and discussed, as well as how they have been used; underlying information of how they work will be also shown. The overall system shown in Figure 1 is made of a robotic hand, a BCI headset and an OpenViBE Software interface combined with an open source embedded platform, namely an Arduino board.

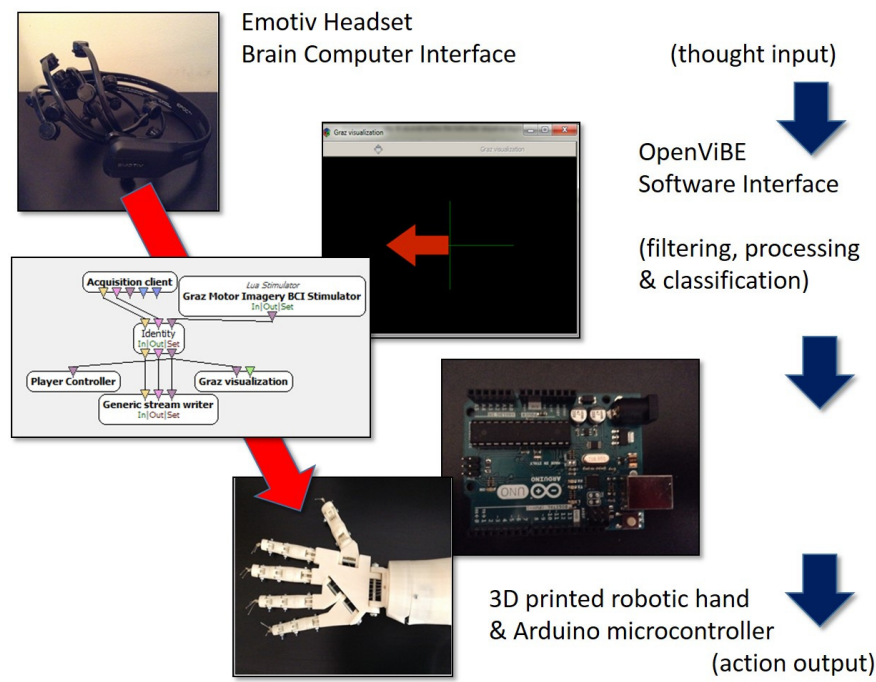

Figure 1 - Overall Design of Framework

Figure 2 displays the assembled 3D printed prosthetic hand that will be used in this research project, as current prosthesis can be very expensive and by 3D printing a much cheaper but still functional version allowing for customisable prototypes to a user's specific design quickly and affordably. The hand design has been inherited from Thingiverse database (2012): STL files have been then imported within an environmental software of HP 3D design and manufactured in ABS (Acrylonitrile Butadiene Styrene) materials by means of a HP Designjet 3D printer.

Concerning the BCI, the chosen headset is reported in Figure 3, whereas a brief overview of its cognitive software interface is detailed in par. 2.1. Finally, the architecture of the OpenViBE medium and of the Arduino interface are shown in par. 2.2 and 2.3, respectively, where the main components of the project will be introduced.

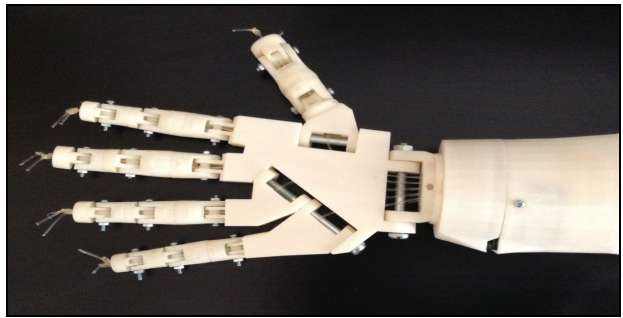

Figure 2 – The 3D Printed 5 d.o.f. Prosthetic Hand (design from Thingiverse database, 2012).

## 2.1. EEG Headset & Cognitive Suite

The BCI headset is an Emotiv EPOC+ premium, i.e. a commercial scientific contextual EEG headset that will be used to record the EEG real-time signals from the human brain while subjects performing mental tasks (Figure 3). Compared to other medical grade headsets the number of electrodes of this device is limited to 14 ones; however in the commercial market this device is reasonably a good compromise between efficiency and economical cost.

The EEG headset is limited to the following 14 electrodes as shown in Figure 4, namely the electrodes *AF3*, *F7*, *F3*, *FC5*, *T7*, *P7*, *O1*, *O2*, *P8*, *T8*, *FC6*, *F4*, *F8*, and *AF4*. Along with this device is the Research Edition (Software Development Kit) SDK that will be used by the used software to acquire the EEG signal data from the headset.

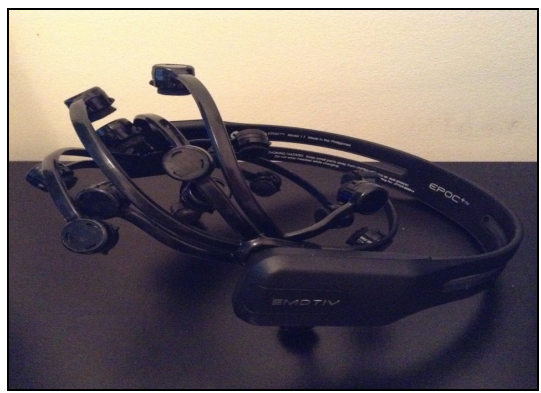

Figure 3 – the BCI Emotiv EPOC+ Headset.

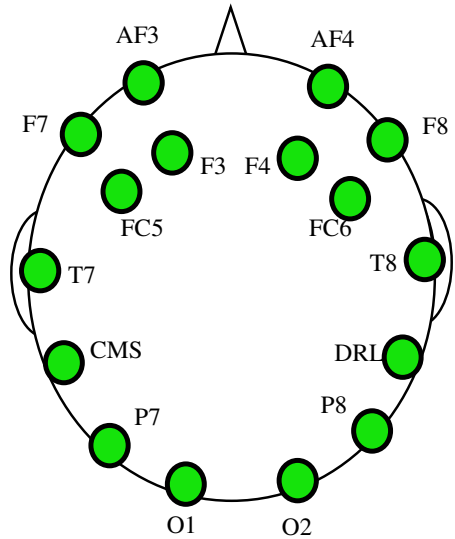

Figure 4 – The Emotiv EPOC+ Electrodes.

For the 1st proposed cognitive Framework is the Emotiv software Cognitive Suite, which will be used to acquire the brain signals, train a the classifier in order to properly cluster the human thoughts into the class and finally control the prosthesis accordingly. The Cognitive Suite is shown in Figure 5: a Graphical User Interface (GUI) performs a floating cube that can be manipulated and manoeuvred into different directions by means of the brain signals. A detailed explanation of its implementation is reported in the next par. 3.1.

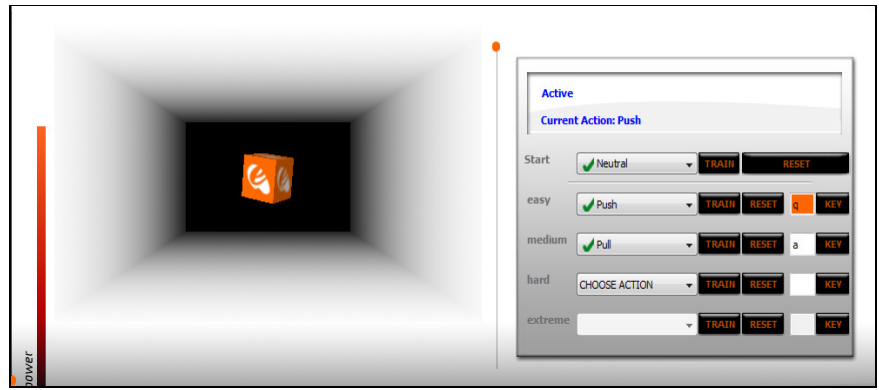

Figure 5 – Snapshot of the Emotiv Cognitive Suite (i.e. the 1st framework).

## 2.2. OpenViBE

The 2nd Framework is developed on OpenViBE software platform (Renard, Y. et al., 2010), due to it being an open source, stand-alone software that can be used for quick and robust prototyping of BCI systems. Figure 6 shows the designer that will be used to develop scenarios that will process, train and classify the EEG signals live from the headset or from previously recorded EEG files. Par. 3.1 will cover the implementation in detail.

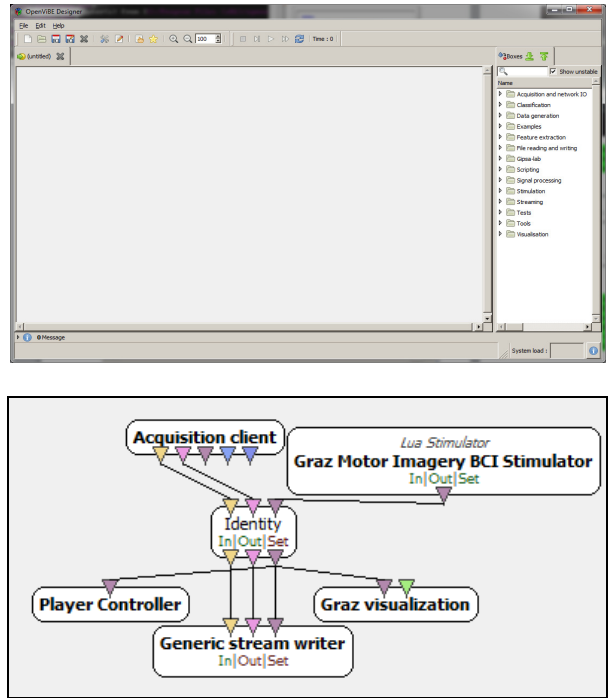

Figure 6 – The OpenViBE main design mask (top panel) and acquisition scenario (bottom panel).

### 2.3. The Arduino's interface

In order to control the prosthetic and robotic hand, an inter-medium interface is needed to feed the output of the cognitive software (namely the $1^{st}$ and $2^{nd}$ frameworks) into the robotic hand. To this aim, an open source and low-cost multi-purpose bard has been chosen: the Arduino UNO board is shown in Figure 7; this microcontroller will be able to receive inputs from a Serial Monitor Window connected to the framework and finally send a driving signal to the set of servomotors within the robotic hand in order to perform an action, either open or close a finger or the whole hand. Such a device has been chosen due to being open-source and is basically a micro-computer that can be used in control type projects and is suitable for both frameworks. The software for Arduino is completely open-source and is an Integrated Development Environment (IDE). The Arduino IDE can provide an easy way for writing code and prototyping. Depending on the adopted framework, different code will be used during implementation.

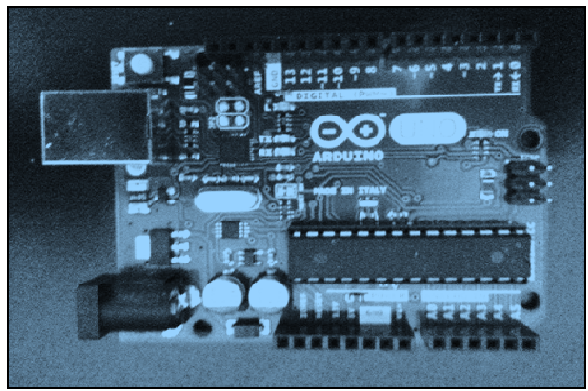

Figure 7 - The Arduino UNO Board.

International Journal of Information Technology Convergence and Services (IJITCS) Vol.6, No.1,February 2016## 3. IMPLEMENTATION
### 3.1. 1st Framework

The Cognitive Suit allows the user to perform the training of different brain activity or thoughts, each of one can then affect a different output. One peculiar feature of the cognitive suite is that it will output the assigned key to whichever window is in focus; for instance in Figure 5 the cube is being pushed and therefore an assigned key '*q*' is being broadcast (see the text which is highlighted in red colour within the figure). As a consequence, an Arduino serial monitor can easily capture the appropriate string letter and decode the message into a command to the prosthetic device performing the finger or hand motion.

A training GUI allowing the user to record the so called 'neutral status' (namely a status in which the cognitive interface will not pursue any output) simply requires the user to relax. For additional action such as push or pull actions, different mental states are requires: for example thinking of numerical problems or imagining limb movements will trigger diverse behaviours and actions accordingly. Due to these different mental states, different skilled users may be able to train a larger number of actions, giving more flexibility in the Arduino code and therefore the individual movements of the robotic fingers.

According to this set-up, an Arduino code was developed which inherits the outcomes of the 1st framework cognitive interface by actuating the hand: Figure 8 shows the mapping code used in the 1st framework that converts 4 keys, namely the '*q*', '*a*', '*w*' and '*s*' status to a different posture of the hand: closed hand configuration, fully open hand configuration, closed thumb and open thumb, respectively. The code also allows to obtain a fully open and close configuration of the hand through the pressure of two more keys.

```
39  if (Serial.available()) {
40      input = Serial.read();
41      if (input == 'q') {      // Close whole hand
42          closePos = 180;
43          one.write(closePos);
44          two.write(closePos);
45          three.write(closePos);
46          four.write(closePos);
47          five.write(closePos);
48      }
49      if (input == 'a') {      // Open whole hand
50          openPos = 0;
51          one.write(openPos);
52          two.write(openPos);
53          three.write(openPos);
54          four.write(openPos);
55          five.write(openPos);
56      }
57      if (input == 'w') {      // Close one
58          onePos = 180;
59          one.write(onePos);
60      }
61      if (input == 's') {      // Open one
62          onePos = 0;
63          one.write(onePos);
64      }
```

Figure 8 – The movement mapping between the Cognitive Suite output and the postural robotic hand configuration.

29

## 3.2. 2nd Framework

The 2nd cognitive framework is designed by using the OpenViBE Acquisition Sever and the Emotiv Research SDK: this latter one acquires the incoming EEG signals and sends them to the OpenViBE platform.

The acquisition scenario (Figure 6, bottom panel) displays to the user a GUI that every 4 seconds will go through the process of a black background, followed by a green cross to indicate a new cue is going to appear and finally the cue in the form of a red arrow that points either left or right (Figure 11). When the cue appears the user must imagine movement of that particular direction until the next green cross appears. This ensures that there will be enough samples per cue for the duration of the acquisition. The more sample the system has for further classification, the better the result will be for accuracy. This stage resumes the training for both the BCI and the user. The longer this scenario is explored, the better the performance of the user from practice and the better the result the BCI will acquire for further processing.

The acquired EEG signals are then used by a further phase in order to train a Common Spatial Pattern (CSP) filter, which was firstly used to improve the separation of two types of input signals, namely the left and right motor imagery. For this purpose, the CSP filter successfully removes the noise and artefacts which are inherently present within these data. The CSP special filter algorithm is applies a linear transformation to project the EEG data in a low-dimensional spatial space with a projection matrix. The rows of this matrix are the weights for channels and this transformation can maximize the variance of two class signal matrices. Let's assume that $c_l$ and $c_r$ are the pre-processed EEG matrices under the two classes (i.e. the left and right imagery), with $N \cdot S$, where $N$ is the number of data channels and S is the number of samples per each channel. Therefore the normalized spatial covariance is computed with the equations (1) where $c^s$ is the transpose of $c$ and trace ($E$) computes the sum of the diagonal elements of $E$. It holds:

$$c_l = \frac{c_l \cdot c_l^s}{trace(c_l \cdot c_l^s)} \qquad c_r = \frac{c_r \cdot c_r^s}{trace(c_r \cdot c_r^s)} \qquad (1)$$

After computing the spatial covariance, the averaged normalized covariance $ANC_l$ and $ANC_r$ are calculated by averaging all the trials of each group. Now the composite spatial covariance can be computed with the equation (2) where $m_0$ is a matrix of eigenvectors and $\sum$ is the diagonal matrix of eigenvalues.

$$ANC = ANC_l + ANC_r = m_0 \sum m_0^s \qquad (2)$$

Figure 9 summarize the application of the CSP spatial filter: namely, the filter is offline trained in order to remove the higher frequencies that are not informative with frequency response between 8 and 30 Hz of the motor imagery task. This processing also discriminates the remaining signals into the left and right classes and four blocks. Once the processing has completed, a configuration file is returned: this set-up can be furtherly used in the next processing scenarios to finally perform the taught filtering task.

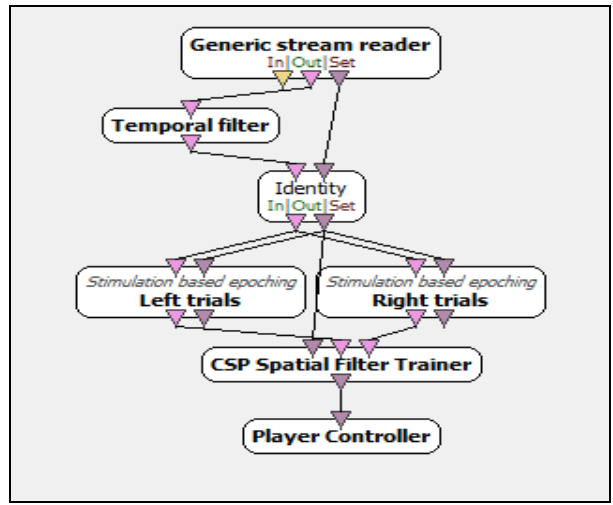

Figure 9 – Signal Filtering within the OpenViBE architecture.

In order to finally classify the signals, Lotte et al. (2007) investigated the classification algorithms of BCI systems and discussed how linear classifiers "are probably the most popular algorithms for BCI applications". Therefore a Linear Discriminant Analysis (LDA) is then applied to the signals: this latter one will "use hyperplanes to separate the data representing the different classes" (Lotte et al., 2007); two different inputs are expected: a negative class and a positive class, for instance *C(l|r)=-1* and *C(l|r)=+1*. Figure 10 refers to the implemented set-up in order to train the classifier and distinguish between the left and right imagery movements.

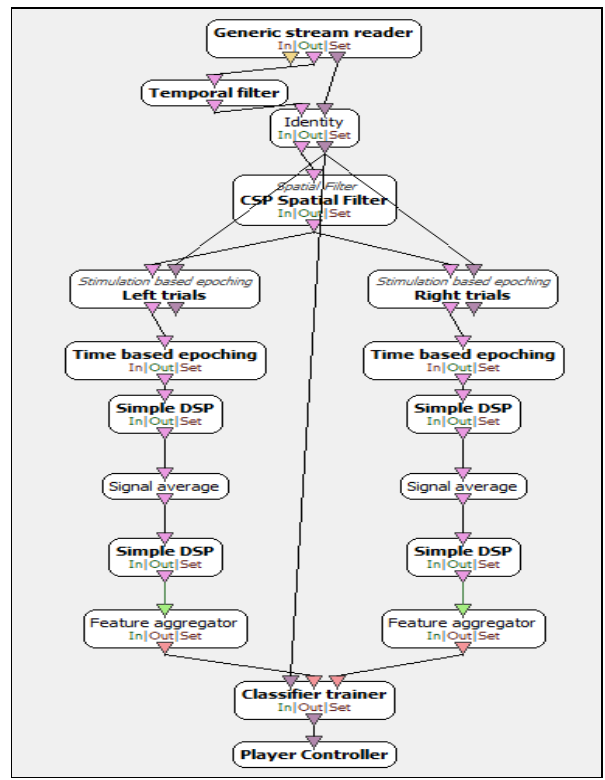

Figure 10 – Signal Classification within the OpenViBE architecture.

International Journal of Information Technology Convergence and Services (IJITCS) Vol.6, No.1,February 2016

After the training analysis of the signal processing, the thought classification is performed: at this stage the software will present the user with a GUI (Figure 11) that allows for the examining of the previous training steps by replaying the acquisition scenario (top panel of Figure 11) with added feedback of the user's thoughts (bottom panel of Figure 12) as well as a confusion matrix in order to display the *instruction flow* and the *classification flow*. The instruction flow is received from the acquired data (i.e. the desired class) whereas the classification flow is the effective output of the classifier processor (i.e. the obtained class). An accuracy measure is also calculated and displayed, which refers to the real time classifier by comparing the inputs from the result of the classifier processor with the targets received from the stimulation filter.

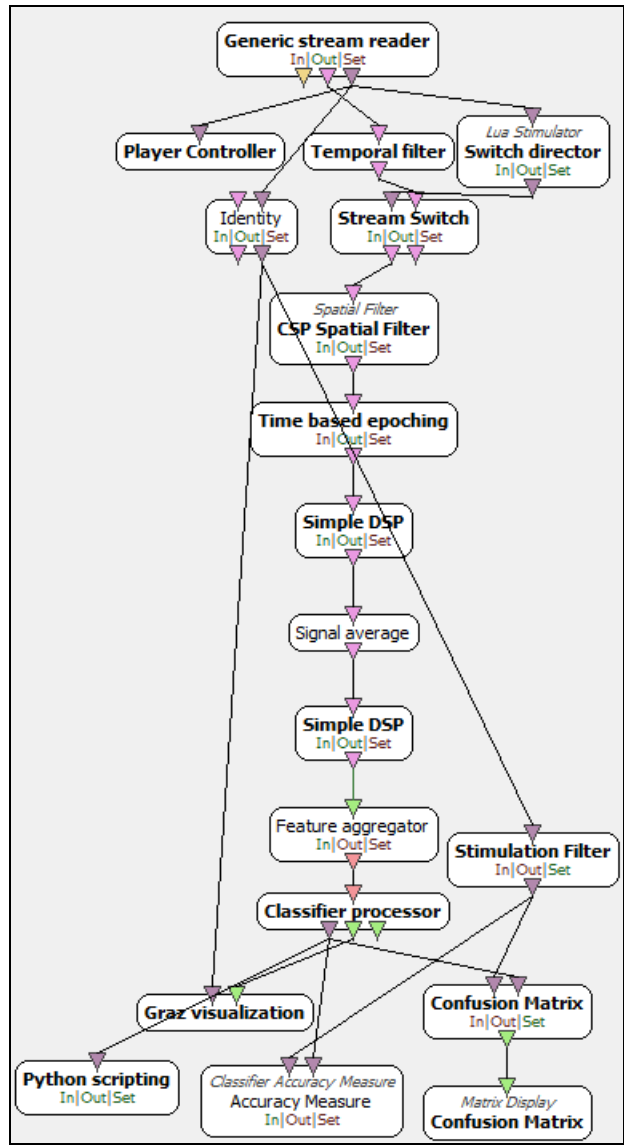

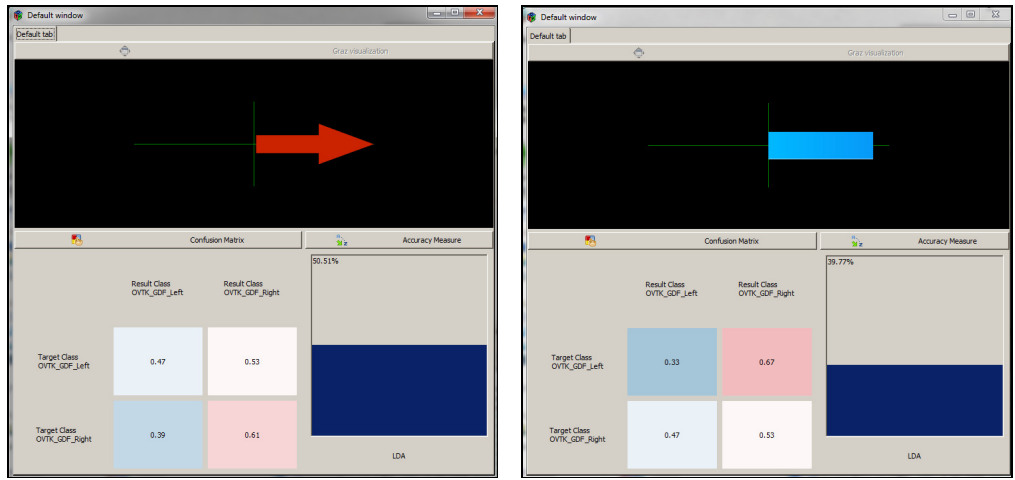

Figure 12 – Visualization of the OpenViBE designed GUI for thought training (top panel) and of the Direction of the Motor Imagery Strength after the training (bottom panel).

An Arduino code analogous to the one which has been reported in Figure 8 is also implemented and used within the framework in order to then activate the robotic hand: finally, when either left or right motor imagery is detected, the whole hand will open or close.

## 4. RESULTS

Results of the 1st framework demonstrated successful control of an external device, namely the 3D arm and the Arduino UNO, by using the EEG signals. In this it was noticed that when the number of actions trained is increased – namely a set of 8 trials -, the accuracy of the system to distinguish between them decreased as shown in Table 1. This issue is likely caused by the processing techniques in the background and as a result means that training can be hard as interference such as facial movements such as but not limited to; blinking, eyebrow movements, clenching of the teeth or even swallowing as well as a limit to know how many movements a user can make.

Table 1 - Emotiv Cognitive Suite Results (i.e. the 1st framework)

| Action  | Assigned Key | Accuracy (%) |
|---------|--------------|--------------|
| *Neutral* | N/A          | 100          |
| *Push*    | q            | 85           |
| *Pull*    | a            | 80           |

The lack of control over the processing in the background for instance the choice in EEG method, the filtration and the classifier makes this system less desirable regardless of it successful functionality. A prosthetic would need to robust and reliably for real world use and a system that cannot allow the user to have a consistent control would not be of use.

Therefore the second framework is designed in order to overcome such drawbacks. In this framework, depending on the scenario and the motor imagery focus, the accuracy percentages could vary. Table 2 shows that on the offline analysis the system could achieve an accuracy of 79.92%, whereas the accuracy slightly decreases at 74% in the real time conditions. Although this is a suitable high accuracy, further testing and analysis showed that the accuracy of the left motor imagery vs. the right one were different: a 100% accuracy could be achieved for the left motor imagery, while the right one could only achieve a value of 56.58%. As the system can properly perform on the left motor imagery, a better training maybe performed for the right motor imagery or this lower performance could be related to the inter subject variability and be improved by recruiting more subjects.

Table 2 - OpenViBE Scenario Results (i.e. the 2$^{nd}$ framework)

| Scenario | Motor Imagery | Number of Cues | Accuracy (%) |
|---|---|---|---|
| *Experiment Analysis Scenario Performance* | Left & Right | 40 | 79.92 |
| *Real-time Two Overall Performance* | Left & Right | 40 | 74 |
| *Real-time Two First Test* | Left | 40 | 100 |
| *Real-time Two Second Test* | Right | 40 | 56.58 |

## 5. CONCLUSIONS

This work presented the design and implementation of an EEG based BCI using motor imagery: two frameworks were presented, which performed a superior version that filter and classify the EEG signal before they output a message to Arduino to control the prosthetic. As a result this study has successfully achieved all of the aims to create a fully functional system and proven the feasibility of the current design systems for highly accurate classification with two robust but affordable systems that could be applied patients requiring applications for external control. Finally, a BCI system takes the raw EEG signals and processes them into a viable source of controlling devices through the OpenViBE platform. In the future, other approached may be explored (Secco et al, 2002; Matrone et al, 2009; Secco et al, 2001; Magenes et al, 2008), as well as another technical methodology that provides a better transference of classifier labels or a box support for serial communication could be implemented rather than using the proposed bridge between the different software platforms.

The 3D printed hand design could be improved slightly, so that the Arduino board can be placed inside of the arm for portability rather than the current setup with wires to a breadboard and then to the Arduino. This change would require moving all of the servos into different positions throughout the arm instead of all together and would have to be considered so as to not affect the functionality of the arm. At the time of writing this paper a slightly newer version of that incorporated the majority of these design changes by creating a larger sized arm than the one used in this study can be seen in the new Thingiverse database (2012).

## ACKNOWLEDGEMENTS

We thank you all staff of the Department of Mathematics and Computer Science for their valuable support and in particular: Mr. Rory McDaid, Mr. Mark Barrett-Baxendale, Dr. David Reid and Prof. Atulya Nagar.

This work was presented in thesis form in fulfilment of the requirements for the MSC in Computer Science for the student D. Elstob under the supervision of E.L. Secco from the Robotics Laboratory, Department of Mathematics & Computer Science, Liverpool Hope University.

## Authors

Daniel Elstob, born in 1992, graduated in Computer Science in 2014 and he is finalizing a Masters of Computer Science in 2016. His main interest is on brain-computer interfaces and recognition systems.

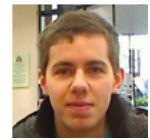

Emanuele Lindo Secco, born in 1971, graduated in Mechanical Engineering in 1998 and received a PhD in Bio-Engineering & Medical Computer Science in 2001. From 2003 to 2014, he has been working for diverse Institutions and Research Centres (Rehabilitation Institute of Chicago, USA; University of Bologna, Italy; European Centre for Training and Research in Earthquake Engineering, Italy; King's College London, UK). In 2015 he joined the Department of Mathematics and Computer Science, Liverpool Hope University as Lecturer. Dr. Secco has been working on neural network controllers, bio-mimetic robots & systems, wearable sensors. His main interest is on human-like robotics.

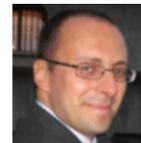